# Efficient parallel strategy for molecular plasmonics – a numerical tool for integrating Maxwell-Schrödinger equations in three dimensions

Maxim Sukharev[1,2]

[1]College of Integrative Sciences and Arts, Arizona State University, Mesa, Arizona 85212, USA
[2]Department of Physics, Arizona State University, Tempe, Arizona 85287, USA

**Abstract**

An efficient parallelization approach to simulate optical properties of ensembles of quantum emitters in realistic electromagnetic environments is considered. It relies on balancing computing load of utilized processors and is built into three-dimensional domain decomposition methodology implemented for numerical integration of Maxwell's equations. The approach employed enables directly accessing dynamics of collective effects as the number of molecules in simulations can be drastically increased. Numerical experiments measuring speedup factors demonstrate the efficiency of the proposed methodology. As an example, we consider dynamics of nearly 700,000 diatomic molecules with ro-vibrational degrees of freedom explicitly accounted for coupled to electromagnetic radiation crafted by periodic arrays of split-ring resonators and triangular nanoholes. As an application of the approach, dissociation dynamics under strong coupling conditions is scrutinized. It is demonstrated that the dissociation rates are significantly affected near polaritonic frequencies.

**Introduction**

Plasmonics has been attracting appreciable interest for many years due to a variety of intriguing applications in chemistry [1], engineering [2], and biology [3]. Remarkable progress in nanofabrication in recent years [4] has allowed researchers to dwell into true nanoscale fabricating plasmonic systems with a riveting spatial precision [5]. The basis for various applications of plasmonic materials is their ability to sustain surface plasmon-polariton resonances (SPPRs) [6] – an optical phenomenon due to collective oscillations of conduction electrons resulting in evanescent electromagnetic (EM) modes residing on a metal/dielectric interface in very small (far below diffraction limit [7]!) volumes [8]. Great sensitivity of SPPRs to surface roughness, nanoparticles' shapes, incident field polarization is relied on to create artificial optical materials with desired properties [9]. Typical resonant wavelengths cover nearly entire visible part of the spectrum and are being pushed into deep infrared [10].

The strong EM field localization at SPPR frequencies recently opened a new research direction, namely polaritonic chemistry [11]. By utilizing plasmonic systems as resonant cavities one can investigate how optics of quantum emitters (such as molecular aggregates, quantum dots, transition metal dichalcogenide monolayers, etc.) changes. When the coupling strength between ensembles of quantum emitters and a local EM field surpasses all the damping rates, the system enters the strong coupling regime forming polaritonic states [12], which have properties of both light and matter. In addition to various fundamental questions arising from modeling, it has been shown that such materials can lead to modified chemistry [13]. Furthermore, it has become a real playground for theorists [14]. The primary source of their wonder is to build self-consistent multiscale models that merge semi-classical macroscopic world with true quantum nature of matter at the nanoscale [15]. Indeed, when spatial scales shrink to nanometers one begins to wonder whether classical electrodynamics merged with quantum dynamics describing quantum emitters can be used as a quantitative tool [16]. In that sense experiments are currently well ahead of the theory.

Quantum mechanics dominates molecular structure and dynamics, and even the mostly classical motions of heavy atomic nuclei show significant quantum effects due to their interaction with the molecular electrons. Similarly, light-matter interaction which is often treated classically can have strong quantum ramifications [17]. As quantum technology is evolving from concept to practice, a better understanding of the level at which such phenomena should be described is needed for two reasons. First, quantum coherence, entanglement, interference, and collective response are at the heart of processes that dominate formation, measurement, functionality and control of quantum devices [18]. Second, and central to the proposed effort, computational studies of such processes in realistic many body systems are too costly to describe on a purely quantum level [19]. The latter observation is the reason for ongoing efforts to develop mixed quantum-classical methodologies for describing molecular processes, and much progress has been made in developing such methodologies for the dynamics of both electrons and atomic nuclei in molecular systems [20].

As we advance our understanding of the physics of such systems the need for multiscale simulations arises. The major challenge is to combine Maxwell's equations with quantum dynamics pertaining to quantum emitters driven by EM radiation, which results in a highly unbalanced load of multiple processors when done conventionally. Due to this very reason many models are limited by two-level emitters, for which quantum dynamics is well described by simple Bloch equations [21]. The situation drastically changes when one is in need of considering the internal degrees of freedom of each emitter. As microscopic dynamics becomes more involved it severely slows down calculations when the number of emitters is increased.

Due to apparent complexity of chemical dynamics at the nanoscale interfaces theoretical efforts are primarily focused on developing models either based on nonrelativistic Pauli-Fierz Hamiltonian [22, 23] or phenomenological approaches such as Dicke's model [24], [25] , and Holstein-Tavis-Cummings models [26] and combining those with the density functional theory [27] and molecular dynamics [28]. Usually, these models are explored in a single exciton basis, which puts computational limitations on vibrational dynamics when many molecules are included [11]. The latter, if considered quantum mechanically along with electron dynamics is numerically burdensome. Although the exploitation of the phenomenological models has led to many interesting observations including qualitative description of pump-probe experiments carried out for $W(CO)_6$ molecules in a resonant IR cavity [29], one of the drawbacks of this approach as it is restricted to a single cavity electromagnetic mode (rarely just a few are considered [20]) and does not allow exploring plasmonic cavities to their full potential on the quantitative level.

Nonetheless, the field of research investigating strong light-matter interaction at the nanoscale is as hot as it has ever been with a large amount of terrific original papers published nearly on a monthly basis sometimes sparking very interesting exchanges between theorists [30] and experimentalists [31, 32]. In the past few years, a handful of excellent reviews have appeared summarizing the current state-of-the-art in theory and experiments, providing authors' perspectives on what remains to be discovered [16, 25, 33-44]. In their seminal paper Galego et al. [45] showed that vibrational dynamics of molecules in resonant cavities under strong coupling conditions happens on modified potential energy surfaces "dressed" with a cavity mode. This in turn leads to avoided crossing and creating of local barriers explaining why chemical reactions in optical cavities can be manipulated. It was also demonstrated that under certain conditions conventional Born-Oppenheimer approximation may fail due to strong electron-phonon (i.e., vibration) coupling [46-48]. A wide range of theoretical papers followed this exploring photoisomerization reactions [49], nonadiabatic transitions induced by cavities [50], electron [51] and energy transfer [52, 53], electron-vibrational dynamics of rhodamine molecules [20], nonradiative dynamics in photochromic molecules [54], and vibrational strong coupling between a cavity and a water molecule [55]. We also note recent works by Schäfer that laid down a new efficient approach in treating quantum dynamics of polaritonic systems [56, 57].

Furthermore, it must be noted that significant progress in developing highly parallelized electronic structure packages has been made [58]. In addition to many commercial software there exists a handful of very efficient and relatively easy to use open-source packages, such as CP2K [59], NWChem [60], OCTOPUS [61], and SALMON [62] just to name the most versatile and well-tested ones. In their recent work [63], the researchers from University of Tsukuba continued impressive set of works [64-66] and combined the real-time time dependent density functional simulations with the dynamics of the vector and scalar potentials on one of the largest supercomputers on the planet, namely Fugaku. Simulations have been carried out for more than $10^4$ atoms considering both the electron dynamics and the ionic motion along with self-consistent electromagnetic interactions.

However, the vast majority of theoretical research of polaritonic systems is limited to either a fundamental cavity mode with no **k**-dependence and/or simple Fabry-Pérot geometries. Quantum dynamics of molecules in these cases is usually treated on multiple processors each taking care of individual molecules [28] without really worrying about electrodynamical aspects as most of works are concerned with steady-state solutions of corresponding equations. This evidently puts significant constraints on exploring very rich ro-vibrational dynamics of molecules at complex plasmonic interfaces and a wide variety of phenomena associated with it.

In many experiments in molecular plasmonics geometries studied include complex metal/dielectric/molecules interfaces and are significantly inhomogeneous in space. Such systems can be described via semiclassical models that couple Maxwell's equations with corresponding quantum equations of motion for molecules. A typical size of a three-dimensional grid with molecules may reach tens of millions of points, thus making integration of corresponding Maxwell's equations coupled to molecular dynamics very slow in spatial regions occupied by molecules. Conventional three-dimensional domain decomposition procedure (discussed in detail in the next Section) splits the simulation domain into smaller sub-domains, with each sub-domain being carried out by a single processor. Each processor thus performs a different number of floating-point operations while equations are marched in time depending on which part of the system it describes. This leads to a highly unbalanced load of all processors and, as a consequence, results in highly inefficient processor usage and long execution times.

One can, however, note that adding extra level of parallelization, which distributes computational load for the quantum dynamics equally among all processors and maintaining three-dimensional domain decomposition for Maxwell's equations should lead to faster execution times. This paper discusses in detail such an approach demonstrating its effectiveness and nearly linear speedup factors.

We combine computational electrodynamics carried out via the finite-difference time-domain (FDTD) approach that considers electric and magnetic fields (rather than vector and scalar potentials) with quantum dynamics of a large number of molecules on a supercomputer and apply this method to tackle collective optical effects in plasmonic cavities of various configurations. Anticipated benefits include natural consideration of short pulses since the methodology is implemented in time domain; direct access to various transients on different time scales including chemical dynamics; strong coupling effects induced by chiral plasmonics. At the heart of our method lies an efficient parallelization strategy that combines a three-dimensional domain decomposition parallelization procedure along with a proper balancing of the computational load of processors utilized in simulations. The paper organized as follows: first, we review conventional domain decomposition methods for integration of Maxwell's equations and demonstrate their great scalability; next, we introduce quantum emitters as a part of simulations and describe our new methodology for efficient parallel simulations; finally, we utilize proposed method to analyze quantum dynamics of molecules in a plasmonic cavity under strong coupling conditions and show that dissociation rates are significantly affected by polaritonic states.

**The domain decomposition methodology for FDTD**

Classical description of EM radiation in plasmonics has been proven to be a very reliable quantitative tool [67]. Its foundation is based on numerical integration of Maxwell's equations written in a differential form

$$\frac{\partial \mathbf{B}}{\partial t} = -\nabla \times \mathbf{E},$$
$$\frac{\partial \mathbf{E}}{\partial t} = c^2 \nabla \times \mathbf{B} - \frac{1}{\varepsilon_0} \mathbf{J}, \quad (1)$$

where $\mathbf{E}$ and $\mathbf{B}$ are electric and magnetic fields, respectively, and the current density has the form $\mathbf{J} = \frac{\partial \mathbf{P}}{\partial t}$ with $\mathbf{P}$ being the macroscopic polarization that follows equations of motion describing materials' response to EM excitation.

Equations (1) are discretized in space and marched in time via the leap-frog propagation scheme following the FDTD method [68]. Spatial discretization is accomplished by employing finite differences utilizing central difference approximation of spatial derivatives. EM field components are also spatially positioned in accordance with the Yee cell [69]. Conventional approach to propagate resulting equations in parallel on multiple processors follows [70] one of the three methodologies schematically depicted in Fig. 1, namely, 1D (slab) decomposition (the entire computational domain is split into 4 parallel slab-type subdomains), 2D (pencil) decomposition (the domain is split into 16 pencil-type subdomains), and 3D decomposition

(splitting the domain into 64 subdomains) [71]. Here each subdomain is carried by a single processor. Due to the curl nature of the Maxwell equations each decomposition scheme requires sending and receiving data on the boundaries between neighboring domains. In this paper we utilize the 3D domain decomposition implementing sending and receiving operations on each of six faces of each subdomain as shown in Fig. 1d. Such an approach results in great scalability as demonstrated below.

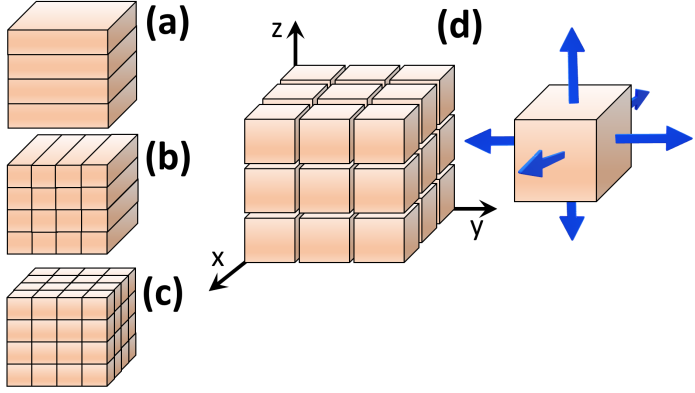

Fig. 1. Domain decomposition strategies. Panels show: (a) 1D (slab) decomposition, (b) 2D (pencil) decomposition, and (c) 3D domain decomposition as implemented in this paper. Panel (d) schematically depicts 3D domain decomposition and the necessity of sending and receiving data on all six faces of each subdomain indicated as blue arrows.

To demonstrate the effectiveness of the 3D domain decomposition scheme we consider a periodic array of Au split-ring resonators (SRRs) [72] shown in Fig. 2a. The array is placed on a semi-infinite glass substrate with a refractive index of 1.52 and is covered with a 60 nm thick polyvinyl alcohol (PVA), which is a semicrystalline polymer often used in experimental plasmonics as a dielectric spacer. The refractive index of the PVA layer is 1.47. Periodic boundaries are set along $X$ and $Y$ and absorbing boundary conditions are implemented along $Z$ to simulate an open system. We implement conventional highly efficient absorbing boundaries, namely convolutional perfectly matched layers (CPML) [73]. The array is excited by a plane wave from the PVA side, and transmission, reflection, and absorption are calculated as functions of the incident frequency using a short-pulse technique. The latter allows to obtain a desired spectral response within a single FDTD run [21]. A short pulse with a duration of 1 fs excites the system and EM field is evaluated at each point along detection planes as a function of time. Once the time propagation is over detected field components are Fourier transformed and the outgoing EM energy flux is evaluated. We also average the resulting flux over the detection planes. The material response of metal follows the Drude model resulting in the equation of motion for the current density, $\mathbf{J}$, [74]

$$\frac{\partial \mathbf{J}}{\partial t} + \gamma \mathbf{J} = \varepsilon_0 \Omega_P^2 \mathbf{E}, \qquad (2)$$

where $\gamma$ and $\Omega_P$ are the damping constant and the plasma frequency, respectively. In the calculations below we use the following parameters that describe Au: $\gamma = 0.073$ eV and $\Omega_P = 9.025$ eV [75]. The dimensions of SRRs are shown in Fig. 2b. Calculated absorption spectrum presented in Fig. 2c exhibits two resonances: localized SPPR near 1.5 eV and the magnetic dipole resonance at 0.73 eV. Local EM field distribution corresponding to the resonance at higher frequency tends to be strongly spatially localized in the PVA, while the magnetic mode has higher mode volume.

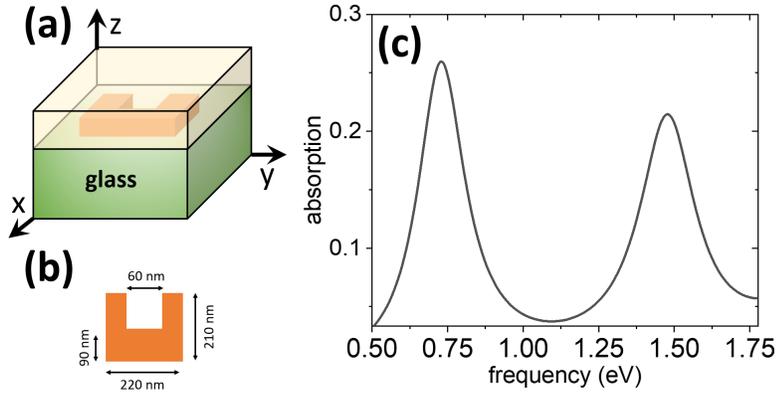

Fig. 2. Periodic array of split-ring resonators. Panel (a) depicts a schematic setup of the geometry. Panel (b) shows dimensions of SRR. The thickness of each SRR is 40 nm and the periodicity along X and Y directions is 360 nm. Panel (c) shows absorption as a function of the incident frequency.

To perform scalability tests using varying a number of processors we consider a domain of the size: 240 (*X*) by 240 (*Y*) by 1243 (*Z*) points resulting in over 70 million grid points. The spatial discretization corresponds to a step size of $\Delta x = 1.5$ nm and the time step is set at $\Delta t = \frac{\Delta x}{2c}$ to ensure numerical stability. Total propagation time to obtain converged spectral response is 300 fs leading to a total number of iterations of $1.2 \times 10^5$. The parallelization of our codes is done via utilizing message passing interface (MPI) subroutines [76]. The 3D domain decomposition can be written in a straightforward manner using MPI_CART subroutines that help to define a given 3D domain decomposition in Cartesian coordinates.

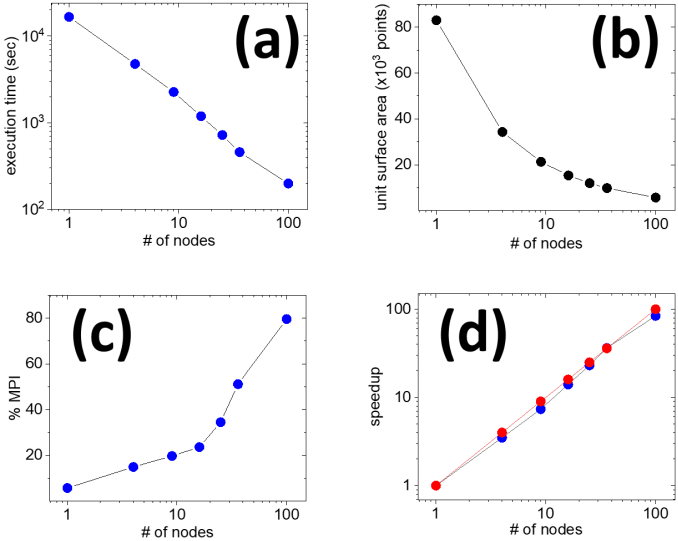

Fig. 3. Scalability of the FDTD code using 3D domain decomposition. Panel (a) shows total execution time as a function of the number of nodes on a double logarithmic scale. Panel (b) shows the surface area (in a number of grid points) of a subdomain as a function of the number of nodes. Panel (c) shows the percentage of the total execution time the code spends on MPI subroutines as a function of the number of nodes. Panel (d) shows the measured (blue circles) and ideal (red circles) speedup factors as functions of the number of nodes on a double logarithmic scale.

In the numerical experiments we utilize a Cray XC40/50 system deployed at ERDC supercomputing resource center named Onyx. All simulations are executed and profiled on nodes equipped with dual Intel E5-2699v4 Broadwell processors having a total of 44 computing cores per node and Cray Aries as interconnects between the nodes. The scaling is done via running the codes on 1, 4, 9, 16, 25, 36, and 100 nodes. Measurements of execution times and a number of MPI calls are performed using Craypat profiling tool. Our choice of the simulation domain above was in part dictated by the fact that proper scaling experiments must be performed on the same physical system with identical dimensions. Additionally, one of the main characteristics needed to be evaluated in such simulations was a speedup factor, i.e., how fast a parallel code performs on $N$ nodes compared to a single node. Ideally one would measure the speedup by executing simulations on a single processor rather than a node. However, this is not feasible for the current setup as it requires too much memory and results in very long execution times. Interestingly, as we show in the next section when a large number of molecules is added to simulations without proper parallelization conventional approach fails even on 1 and 4 nodes due to unacceptably high memory demands.

The main results of this section are shown in Fig. 3. Clearly total execution time scales nearly linearly on the double logarithmic scale with the number of nodes utilized (Fig. 3a) confirming the numerical efficiency of the 3D domain decomposition approach. As a number of nodes increases the surface area of each subdomain (a unit cube carried by a single processor, see Fig. 1d) decreases as shown in Fig. 3b. However, the number of subdomains (equivalent to the number of processors used) increases, which in turn affects the percentage of the total execution time the code spends on serial calculations. This is also demonstrated in Fig. 3c showing the increase in the percentage of MPI calls. The true measure of the parallel efficiency is a speedup fact, i.e., the ratio of the execution time a code spends on 1 node and the execution time it needs to complete simulations on $N$ nodes. Ideally this factor equals $N$, but due to a latency of node-to-node connections and a current load of the network when codes are executed (when other users running their simulations on nearby nodes) the speedup can be drastically affected. Eventually measured speedup saturates as simulations spend more time sending and receiving data between subdomains rather than advancing Maxwell's equations in time. The measured speedup factor and its ideal values are plotted in Fig. 3d. One can see that the measured speed follows closely its ideal values and begins to exhibit a sign of saturation when 100 nodes are used. We also note that the execution time being nearly 5 hours on a single node (44 cores) becomes just 3 minutes on 100 nodes (4,400 cores). It should be emphasized that the definition of the speedup utilized (i.e., speedup relative to execution times on a single node rather than on a single core) could also contribute to nearly ideal values in Fig. 3d. Nonetheless, the efficiency of the 3D domain decomposition is clearly observed.

**Efficient numerical implementation of quantum dynamics of ensembles of molecules coupled via FDTD at plasmonic interfaces**

As a complexity of plasmonic materials increases simulations based on the 3D domain decomposition become less efficient due to unbalanced load of processors, which in turn is because of a significant inhomogeneity in a number of local equations to be propagated. As an example, we consider placing a thin molecular layer on the PVA layer that covers SRRs. Schematically this setup is depicted in the inset of Fig. 4b. Each grid point inside the molecular layer is associated with a set of differential equations taking a local EM field as an input and calculating induced molecular dipole as an output. The latter enters the Ampere law, which in turn changes the total EM field affecting other molecules thus making simulations self-consistent. It is important to note that only a handful of processors need to propagate in time additional equations of motion governing quantum dynamics of molecules. The more complex the molecular model is the higher the unbalance of the load of all processors. We thus are in need to ensure a relatively equal load of processors.

For that purpose, we modify parallel FDTD approach discussed in the previous section in the following way. 3D domain decomposition of Maxwell's equations is kept intact. We pre-calculate a total number of grid points in the molecular layer and spread all molecules equally among all available processors. Each processor advances quantum dynamics on equal footing. In order to merge this approach with the original spatial decomposition of the EM field, we assign local EM field driving a given molecule to a processor that handles dynamics of this molecule. Once the induced molecular dipole is calculated, it is sent back to the 3D decomposition map. Obviously, such an approach can effectively decrease execution times if sending EM field from original grid to all processors and subsequent sending updated dipoles back to FDTD is faster than quantum dynamics. We refer to this method as 3D domain decomposition with molecular mapping.

We employ the same grid used in the previous section and assume that the molecular layer is 18 nm thick. This results in 691,200 grid points inside the layer, each of which is treated as an individual molecule. To demonstrate the effectiveness of the proposed idea we consider molecules described by a two electronic state diatomic model with ro-vibrational degrees of freedom explicitly accounted for. Each molecule follows conventional Born-Oppenheimer approximation that separates electronic and nuclear motion. Corresponding ro-vibrational wavefunctions are propagated on two coupled electronic potential energy surfaces $U_g$ and $U_e$. The rotational degree of freedom is taken into account by expanding nuclear wavefunctions in the basis set of normalized Wigner rotational matrices $D^j_{m,0}$, where $j$ and $m$ are rotational quantum number and its projection on a fixed axis of the laboratory frame.[77]

$$\chi_{g,e}(\vec{R},t) = \sum_{j,m} \xi^{(g,e)}_{jm}(R,t) D^j_{m,0}(\hat{R}). \tag{3}$$

Here $\vec{R}$ denotes internuclear separation vector. In Eq. (3) we adopt the following notation: $\vec{R} \equiv (R, \hat{R})$ denoting the magnitude of the internuclear separation, $R$, and the angular dependence of the molecular axis in the laboratory frame, $\hat{R}$. This expansion is plugged into the Schrödinger equation that subsequently leads to the set of coupled partial differential equations describing ro-vibrational dynamics on potential energy surfaces

$$i\hbar \frac{\partial \xi^{(g)}_{jm}}{\partial t} = \hat{H}^{(g)}_j \xi^{(g)}_{jm} - d_{ge}(R) \sum_{j'm'} \left( E_x X^{j'm'}_{jm} + E_y Y^{j'm'}_{jm} + E_z Z^{j'm'}_{jm} \right) \xi^{(e)}_{j'm'},$$

$$i\hbar \frac{\partial \xi^{(e)}_{jm}}{\partial t} = \hat{H}^{(e)}_j \xi^{(e)}_{jm} - d_{eg}(R) \sum_{j'm'} \left( E_x X^{j'm'}_{jm} + E_y Y^{j'm'}_{jm} + E_z Z^{j'm'}_{jm} \right)^* \xi^{(g)}_{j'm'}, \tag{4}$$

where matrix elements denoted as *X/Y/Z* correspond to the transformation of electric field components $E_{x,y,z}$ from the laboratory frame to the molecular frame[78] and $d_{ge,eg}$ is the transition dipole moment that couples ro-vibrational dynamics on the ground, $U_g$, and excited, $U_e$, electronic states. The effective Hamiltonians in (4) are

$$\hat{H}^{(g,e)}_j = -\frac{\hbar^2}{2\mu} \left( \frac{\partial^2}{\partial R^2} - \frac{j(j+1)}{R^2} \right) + U_{g,e}(R), \tag{5}$$

with $\mu$ being a reduced molecular mass (in numerical experiments below we used the reduced mass of Li$_2$). The induced dipole of each molecule is calculated at each time step. The corresponding macroscopic polarization is evaluated by employing the mean field approximation, which leads to a simple multiplication of local induced dipoles by the molecular number density. Eqs. (4) are propagated numerically in time via split-operator method that employs Fast Fourier Transformation [79].

It is worth emphasizing that the electromagnetic field evaluated in Cartesian coordinates is advanced forward in time in the laboratory frame, when Maxwell's equations are directly coupled to quantum dynamics of molecules using conventional FDTD. In spatial regions occupied by molecules, the local electric field is used to march in time equations of motion describing quantum dynamics in molecular frame (since the electric field varies significantly over space the molecular frame is local and also is spatially variant). The induced dipole moment obtained from the quantum dynamics is then coupled back to the Maxwell equations via polarization current. The coupling from and back to the laboratory frame relies on the ability to analytically integrate the angular part of the quantum dynamics. This, in turn, if done properly gives us *direct access to orientation and alignment dynamics in plasmonic environments*.

The nuclear coordinate $R$ is discretized with a step of 0.092 a.u. between 3 a.u. and 50 a.u. Initial conditions correspond to the ground ro-vibrational state ($j = 0$, $m = 0$) in the ground electronic state. It is assumed that the parity of two electronic states is different thus vertical electronic transitions populate rotational states following the selection rule $\Delta j = \pm 1$. Since molecules are spatially located in the near-field zone of SRRs, local field polarization varies in time and different $m$ states are populated as well. In the numerical experiments discussed below we examine a simple case of a low intense pump such that the rotational basis is truncated at $j = 1$. We also evaluated the effectiveness of the codes by considering rotational wavepackets with different values $j$ as initial conditions (up to $j = 5$) and found that the scalability of the codes with a larger basis of rotational states remains intact.

In the simulations we consider the transition dipole moment of 10 Debye and assume it does not depend on the internuclear coordinate $R$. Since Eqs. (4) are numerically propagated one can use potential energy surfaces in any form, i.e., either analytical or numerical. As an example, we consider the ground potential energy surface in the form of a Morse potential

$$U_g = D_g \left(1 - \exp\left(-\alpha_g \left(R - R_g\right)\right)\right)^2 - D_g, \tag{6}$$

where parameters (all in a.u.) are: $D_g = 3.86 \times 10^{-2}$, $\alpha_g = 0.458$, $R_g = 5.06$. Note that these parameters correspond to Li$_2$ molecule. The excited electronic state is taken in the form of a repulsive potential allowing us to investigate dissociation dynamics

$$V_e(R) = 2D_e \exp\left(-\alpha_e \left(R - R_e\right)\right), \tag{7}$$

with parameters (also all in a.u.): $D_e = 1.26 \times 10^{-2}$, $\alpha_e = 0.364$, and $R_e = 3.98$. The parameters of the excited state are artificially adjusted to make molecules resonant with the localized SPPR mode of the plasmonic array. Both potentials are shown in Fig. 4a.

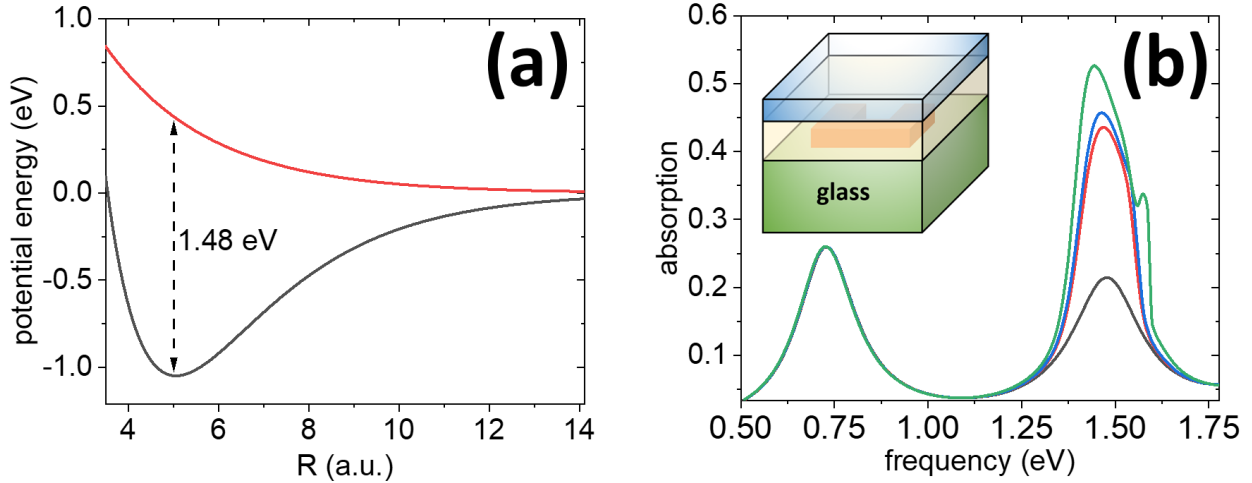

**Fig. 4**. Molecular plasmonics simulations. Panel (a) shows two electronic potential energy surfaces as functions of the internuclear distance. The vertical dashed line shows the value of the vertical gap, which is set to match the localized SPPR of the plasmonic system. The geometry of simulations is depicted in the inset of panel (b), where the molecular layer shown in blue is placed on top of the PVA layer covering SRRs. The main panel (b) shows absorption as a function of the incident frequency without molecules (black) and with molecules at different molecular concentrations: (red) $8\times 10^{25}$ m$^{-3}$, (blue) $10^{26}$ m$^{-3}$, and (green) $2\times 10^{26}$ m$^{-3}$.

Since ro-vibrational dynamics occurs on a significantly longer time scale compared to electron dynamics in plasmonic materials, there is no need in updating molecular dipoles at every time step $\Delta t$ pertaining to the time discretization set in FDTD simulations. From a series of numerical simulations testing convergence, we found that molecular dynamics can be safely evaluated every 40$^{th}$ time step, i.e., every 0.1 fs. The system is probed in the same manner as in the previous section – we consider a linear optical response to a low intense probe. Initial ro-vibrational wavefunction localized in the ground electronic state is partially projected onto the excited state and begins to move down the dissociative slope. The absorption shown in Fig. 4b noticeably increases once molecules are included. With the increasing molecular concentration, the absorption further increases and begins to exhibit two peaks. The latter is a manifestation of the strong coupling between the localized SPP mode and molecules as the coupling strength between the EM mode and ensemble of molecules becomes higher than the damping rates. The resulting two peaks correspond to polaritonic states usually referred to as lower and upper polaritonic states. It is also observed that the shape of the absorption peak becomes clearly not symmetric. This is due to the fact that the Franck-Condon distribution, i.e., the overlap of the initial wavefunction in the ground state with the continuum of states in the dissociative potential, is not perfectly symmetric and the SPP mode is also dispersive.

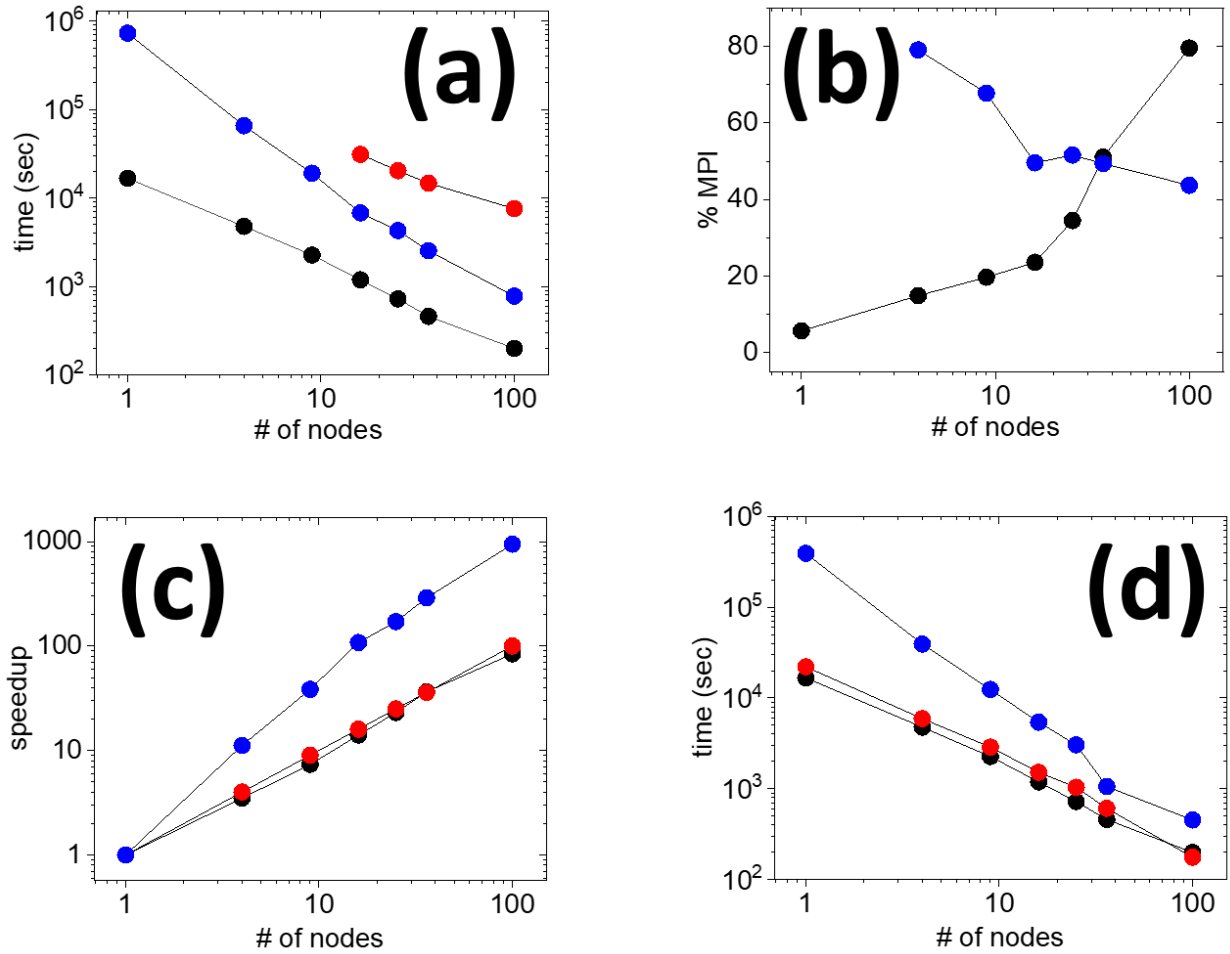

**Fig. 5**. Scaling experiments for molecular plasmonics. Panel (a) shows execution time vs. the number of nodes on a double logarithmic scale for the array of SRRs without molecules (black circles), with mapped molecules (blue circles), with molecules without mapping (red circles). Panel (b) shows the percentage of the total execution time the code without molecules (black circles) and with mapped molecules (blue circles) spends on MPI subroutines as a function of the number of nodes. The speedup factor vs. the number of nodes is shown in panel (c), where the ideal speedup is represented by red circles, black circles show the speedup of the code without molecules, and blue circles show the measured speedup of the code with mapped molecules. Panel (d) shows execution time as a function of the number of nodes for the code without molecules (black circles), without mapping (red circles) and with mapping (blue circles) for molecules with frozen ro-vibrational degrees of freedom.

Our main results are summarized in Fig. 5. Measured execution times for three cases are shown in Fig. 5a. As we noted above when the proposed parallelization method is not employed, and molecules are not equally mapped to all processors, execution times are almost two orders of magnitude longer compared to the case without molecules. Furthermore, the code failed to execute on 9 nodes as the memory needed to be allocated for molecules exceeded available memory. Once the processor load is balanced the executions times scale nearly linearly with the number of nodes on the double logarithmic scale mimicking that obtained in the previous Section. It is informative to examine the percentage of the total execution time simulations spent on MPI calls. Fig. 5b compares the performance of the FDTD code without molecules with the molecular code with mapping. As the number of nodes increases, we see that simulations with mapped molecules spend less time on sending and receiving data via MPI subroutines, but eventually this trend reverses and the execution time on MPI begins to level off. In order to understand this trend, we need to recall how the method with mapped molecules works. At every time step, in addition to send/receive MPI calls required to update EM field via curl equations, updated local electric field components inside the molecular layer are mapped to all processors. As the number of utilized nodes increases, each processor carries a smaller number of molecules thus spending less time on quantum propagation but must also send and receive more data due to molecular mapping. Eventually the time needed for both procedures becomes comparable. This is when the percentage of execution time spent on MPI calls reaches 50% and becomes nearly constant. It is anticipated that for an even larger number of nodes the trend will be reversed and would follow the curve obtained without molecules. Due to highly nonlinear dependance of the execution times on several nodes for the molecular code the measured speedup factor exhibits so-called nonlinear speedup exceeding its ideal values as shown in Fig. 5c. It should be noted, however, that our definition of a speedup was relative to execution times on a single node and not a single processor. Unfortunately, the latter is an impossible task as the memory requirements for a serial code with so many molecules go far beyond accessible memory on Onyx. Thus, the speedup calculated in this manner when exceeding its ideal values should be taken with a grain of salt. Nonetheless, simulations with mapped molecules scale in a similar manner to 3D domain decomposition simulations. What may be a better performance measurement is a speedup of the code with mapped molecules compared to the code with the unbalance load (red circles in Fig. 5a). The ratio of the corresponding execution times on 100 nodes for these two codes reaches 9.74 meaning the code with the balanced processor load runs almost 10 times faster than its unbalanced counterpart on 100 nodes.

Our parallelization strategy suggests that in order to balance the processors load, we equally spread molecules over all processors, which in turn requires completing two sets of sending and receiving operations per time step. The time required to accomplish this may exceed the wait time for simulations with the unbalanced load. Therefore, one needs to estimate whether mapping discussed above is worth the trouble. To demonstrate when such an approach slows down simulations, we consider the same setup as in the inset of Fig. 4b and artificially "freeze" ro-vibrational degrees of freedom. Molecules then become trivial two-level emitters with no internal degrees of freedom. Corresponding equations of motion are reduced to a simple set of Bloch equations that are easily solved using Runge-Kutta $4^{th}$ order scheme. The results of numerical experiments are summarized in Fig. 5d, where we present execution times vs. the number of nodes for the code without mapped molecules (red circles) and with mapping (blue circles). Although the more nodes we use the shorter the time, the execution times for the code without molecular mapping are noticeably shorter. This illustrates the fact that sometimes when a material model is as simple as Bloch equations for a two-level emitter, additional mapping resulting in extra send/receive operations apparently has a negative impact on a code performance. It is thus more efficient to propagate a code with the unbalanced load simply because the number of floating-point operations per molecule is very small.

**Applications: collective dissociation dynamics near resonant plasmonic interfaces**

The main feature of the proposed method is to directly capture collective effects in time domain since we can consider a large number of interacting molecules. To illustrate this, we consider another plasmonic system depicted schematically in the inset of Fig. 6a – periodic array of triangular nanoholes in a thin Au film is placed on a semi-infinite glass substrate and is covered by a thin molecular layer (shown as a blue slab in the inset). The intent is to achieve the strong coupling regime between molecules and one of the SPP modes of the array and investigate the dissociation dynamics under such conditions. Even though a signature of the strong coupling, i.e., so-called Rabi splitting is observed for the periodic array of SRRs (Fig. 4b), periodic arrays of nanoholes in general support higher local field enhancement and thus higher SPP-to-molecule coupling strengths.

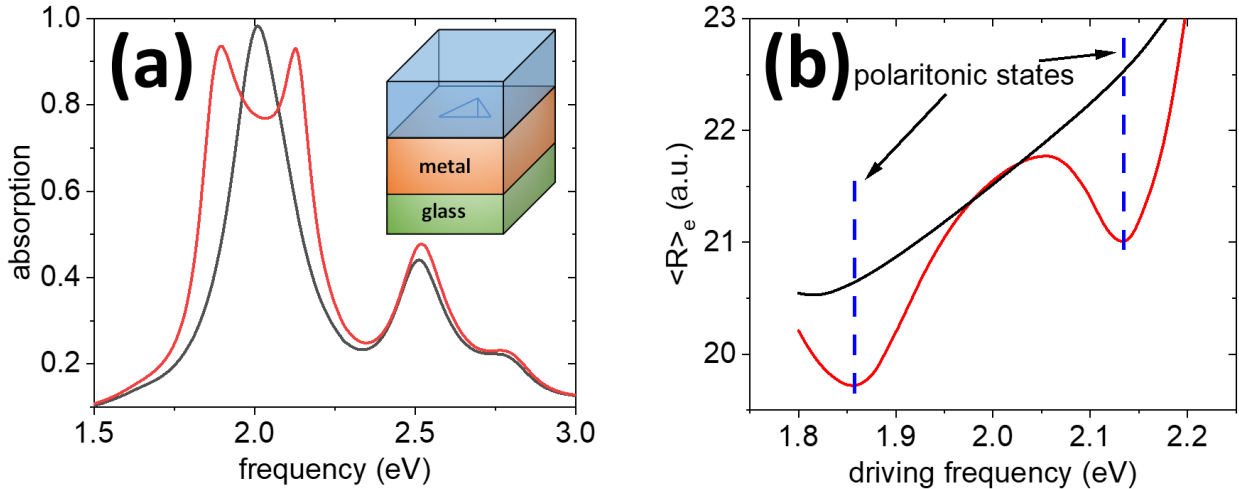

**Fig. 6**. Collective dissociation dynamics at plasmonic interfaces. The inset in panel (a) shows a periodic array of triangular holes in 350 nm thin Au film on a glass substrate (refractive index is 1.52). The system is periodic along X and Y with a period of 350 nm. The triangular hole is in the shape of an equilateral triangle with a side of 230 nm. The molecular layer is placed on top of the metal film and is 18 nm thick. Panel (a) shows absorption as a function of frequency for the array without molecules (black line) and with molecules (red line) with a number density of $2\times10^{26}$ m$^{-3}$. Panel (b) shows an ensemble average internuclear distance calculated using Eq. (8) for molecules in vacuum (black line) and on the metal film (red line). Vertical blue dashed lines show polaritonic frequencies.

We consider the same type of molecules as in previous section making them resonant with a localized SPP mode of the array at 2.01 eV as shown in Fig. 6 by adjusting parameter $D_e$ in Eq. (7). Placing resonant molecules at a moderate molecular concentration on top of the array leads to the splitting of the SPP mode. Such a splitting, referred in the literature as a Rabi splitting, is a unique signature of the strong coupling regime[80] – molecules and the resonant SPP mode exchange energy forming polaritonic states that have properties of both matter and light.[14] To investigate dynamics of a ro-vibrational wavepacket on the dissociative potential energy surface (Fig. 4a, red line) we perform continuous wave (CW) simulations: the system is excited by a weak CW field that is turned on in the first 5 fs and then it continues to drive till the steady-state regime is reached. Initially relaxed molecules are pumped by the local field and undergo vertical transitions from the ground electronic state to the dissociative state. The wavepacket in the excited state begins to move to larger $R$ while being continuously pumped from the ground state.

One may ask several questions regarding the dynamics – how dissociation dynamics varies in and out of the plasmonic system; if there are any collective features of polaritonic states in such dynamics.

To quantify the dissociation dynamics, we calculate the ensemble average internuclear position of the wavepacket, $\chi_e(\mathbf{R},t)$ (from Eq. (3)), in the excited state in the steady-state regime as a function of the CW driving frequency

$$\langle R \rangle_e = \frac{\langle \chi_e | \mathbf{R} | \chi_e \rangle}{\langle \chi_e | \chi_e \rangle}. \tag{8}$$

The results of the simulations are shown in Fig. 6b, where we compare dissociation dynamics near the plasmonic array (red line) with that in vacuum (black line). One can clearly see a striking difference between the two cases. In vacuum the higher the driving frequency is the faster the dissociation becomes, which is easy to understand after examining the excited electronic state (red line in Fig. 4a) – the higher in energy the wavepacket is projected on this state the faster it moves down the potential slope. On the contrary, the dissociation at the resonant plasmonic interface is noticeably slowed down near frequencies corresponding to the polaritonic frequencies of the system (peaks in the absorption spectrum shown in red in Fig. 6a). The physics of this phenomenon is the subject of the separate manuscript [81]. In brief, near polaritonic frequencies the local electric field (being a linear combination of the incident field, scattered field due to the plasmonic array and due to molecules) exhibits enhancement. The ro-vibrational wavepacket, on the other hand, moves on a modified potential energy surface, which is affected by nearby molecules and the resonant SPP field. Such a potential surface becomes shallower at either polaritonic frequency due to the field enhancement and this in turn slows down the dissociation. The observed phenomenon is of a clear collective nature (due to formation of polaritonic states) and cannot be seen in systems with a small number of molecules.

**Conclusion**

We proposed a parallelization approach to simulate optical properties of ensembles of quantum emitters in realistic electromagnetic environments. It relies on balancing computing load of utilized processors and is built into three-dimensional domain decomposition methodology. Our approach enables directly accessing dynamics of collective effects as the number of molecules in simulations can be drastically increased. As an example, we consider dynamics of nearly 700,000 diatomic molecules with ro-vibrational degrees of freedom explicitly accounted for coupled to electromagnetic radiation crafted by periodic arrays of split-ring resonators and triangular nanoholes. Numerical experiments measuring speedup factors demonstrate the efficiency of the proposed methodology. As an application of the approach, dissociation dynamics under strong coupling conditions is scrutinized. It is demonstrated that the dissociation rates are significantly affected near polaritonic frequencies.

**Acknowledgements**

This work is supported by the Air Force Office of Scientific Research under grant No. FA9550-22-1-0175. Computational experiments and scaling simulations are made possible by the Department of Defense through the High-Performance Computing Modernization Program. The author is grateful to Prof. Eric Charron and Prof. Osman Atabek for fruitful and stimulating discussions pertaining to the algebra of angular momentum in quantum mechanics. The author also acknowledges the enormous help by Prof. Eric Charron with providing and explaining parts of his code for quantum wavepacket dynamics. The author is thankful to the reviewer of this manuscript for his/her critical comments pertaining to the parallel methodology employed.